\documentclass[9pt,twocolumn,floatfix,showpacs,pra]{revtex4-1}
\usepackage{amsmath,amssymb,amsthm,mathrsfs,amsfonts,dsfont} 
\usepackage{amsmath,amssymb,amstext}
\usepackage{textcomp}

\usepackage{tikz}
\usepackage{bm}
\usepackage{dcolumn}
\usepackage{booktabs}
\usepackage[scaled]{helvet}
\usepackage{sansmath}
\usepackage{graphicx}
\usepackage{transparent}
\usepackage{color}
\usepackage{gensymb}
\usepackage{multirow}
\usepackage{url}
\usepackage[colorlinks=true]{hyperref}

\begin{document}

\title{Light-matter micro-macro entanglement}

\author{Alexey Tiranov$^{1}$}
\author{Jonathan Lavoie$^{1}$}
\author{Peter C. Strassmann$^{1}$}
\author{Nicolas Sangouard$^{2}$}
\author{Mikael Afzelius$^{1}$}
\author{F\'{e}lix Bussi\`{e}res$^{1}$}
\author{Nicolas Gisin$^{1}$}

\affiliation{$^{1}$Group of Applied Physics, University of Geneva, CH-1211 Geneva 4, Switzerland}
\affiliation{$^{2}$Department of Physics, University of Basel, CH-4056 Basel, Switzerland}

\date{\today}

\newcommand{\ket}[1]{\vert#1\rangle}
\newcommand{\bra}[1]{\langle#1\vert}
\newcommand{\YSO}{Y$_2$SiO$_5$}
\newcommand{\nd}[0]{Nd$^{3+}$:Y$_2$SiO$_5$}
\newcommand{\tranz}{$\pm|3/2\rangle_g \rightarrow \pm|5/2\rangle_e$ }
\newcommand{\trann}{$\pm|5/2\rangle_g \rightarrow \pm|3/2\rangle_e$ }
\newcommand{\trans}{$\pm|1/2\rangle_g \rightarrow \pm|5/2\rangle_e$ }
\newcommand{\transition}{$^4$I$_{9/2} \rightarrow ^4$F$_{3/2}$ }
\newcommand{\X}{\textbf{X}}
\newcommand{\Y}{\textbf{Y}}
\newcommand{\e}{\mathrm{e}}
\newcommand{\mathbbm}[1]{\text{\usefont{U}{bbm}{m}{n}#1}} 
\newcommand{\braket}[2]{\langle#1\vert#2\rangle}
\newcommand{\suppmat}[0]{{Appendix}}


\begin{abstract}
Quantum mechanics predicts microscopic phenomena with undeniable success. Nevertheless, current theoretical and experimental efforts still do not yield conclusive evidence that there is, or not, a fundamental limitation on the possibility to observe quantum phenomena at the macroscopic scale. This question prompted several experimental efforts producing quantum superpositions of large quantum states in light or matter. Here we report on the observation of entanglement between a single photon and an atomic ensemble. 
The certified entanglement stems from a light-matter micro-macro entangled state that involves the superposition of two macroscopically distinguishable solid-state components composed of several tens of atomic excitations. Our approach leverages from quantum memory techniques and could be used in other systems to expand the size of quantum superpositions in matter.
\end{abstract} 

\maketitle
Quantum mechanics has been tested in many situations with a remarkably excellent agreement between theory and experiments. There remains, however, one interesting challenge, namely to demonstrate quantum effects at larger and larger scales~\cite{Brune1996a,Leibfried2005a,Arndt2014a}. This is a timely topic, especially with the advance of quantum technologies that allow one to entangle many kinds of systems involving photons, artificial solid-state atoms, trapped ions, atomic ensembles, nanomechanical oscillators and large molecules, to name but a few. 
These approaches all involve ``individual'' quantum systems (even though each system may be composed of a large number of particles), and should be distinguished from ensemble quantum effects such as superconductivity~\cite{Leggett1980a}. These individual systems offer a unique approach to study macroscopic quantum effects, which raises interesting questions: How far can entanglement hold in such systems? How can one compare different systems?

There are many approaches trying to define what constitute a quantum superposition of macroscopic states~\cite{Frowis2015a,Jeong2015a,Farrow20152a}. The one we use is based on the distinguishability between the states forming the superposition, as formalized in Ref.~\cite{Sekatski2014b}. More precisely, we say that two quantum states are macroscopically distinct if they can be distinguished with a detector that has a coarse-grained resolution, and we use ``macroscopic'' to mean ``macroscopically distinguishable". This introduces some degree of arbitrariness in what should be the minimum level of coarse-graining, which reflects the challenge of defining such a measure. Instead of trying to achieve this, we  use a way to compare different kinds of states to assign them an effective size, as detailed in Ref.~\cite{Sekatski2014a}. Consequently, the number of particles (or photons) is not used to define the macroscopic nature of the superposition state. Rather, the number of particle is a property of the state that, when increasing, makes the two components easier to distinguish with a given coarse-grained detector (and hence look more like distinct macroscopic objects). Interestingly, the more distinguishable the states become, the more challenging it is to experimentally reveal that they have quantum features (such as entanglement in a micro-macro entangled state)~\cite{Sekatski2014b}, which explains why we do not easily observe such kind of states.

Quantum optics offers a powerful approach to study the quantum features of superpositions of macroscopic states. Purely photonic experiments for example have reported on superposition of coherent states with opposite phases~\cite{Ourjoumtsev2006a,Neergaard2006a,Wakui2007a,Ourjoumtsev2007,Vlastakis2013a}, squeezing \cite{Eberle2013a,Beduini2015a} and micro-macro entanglement~\cite{Martini2008,Bruno2013a,Lvovsky2013a,Jeong2014,Morin2014a}. Hybrid systems have also been exploited for micro-macro entanglement where the micro part was an atom and the macro part contained up to 4 photons~\cite{Deleglise2008}.
It was proposed to use mirror-Bose-Einstein condensate to observe macroscopic quantum superpositions between light and matter~\cite{deMartini2010}.
In matter, GHZ-type states have been produced with up to 14 trapped ions~\cite{Monz2011a}. 
Here we report on the observation of entanglement between a single photon and an atomic ensemble containing up to 47 atomic collective excitations, and we give evidence that it constitutes genuine light-matter micro-macro entanglement.

Our implementation, inspired from the proposal of Ref.~\cite{Sekatski2012b} and experiments~\cite{Bruno2013a,Lvovsky2013a}, lies within this scenario. More precisely, we start from two photons entangled in polarization and use a local displacement operation to displace, in optical phase space, one polarization mode of one photon from the pair. The displacement populates one of the polarization modes with a large number of photons, without affecting the amount of entanglement. The displaced photon is then mapped to an atomic ensemble, creating the light-matter micro-macro entangled state.

\begin{figure*}[!t]
\centering
\includegraphics[width=0.95\textwidth]{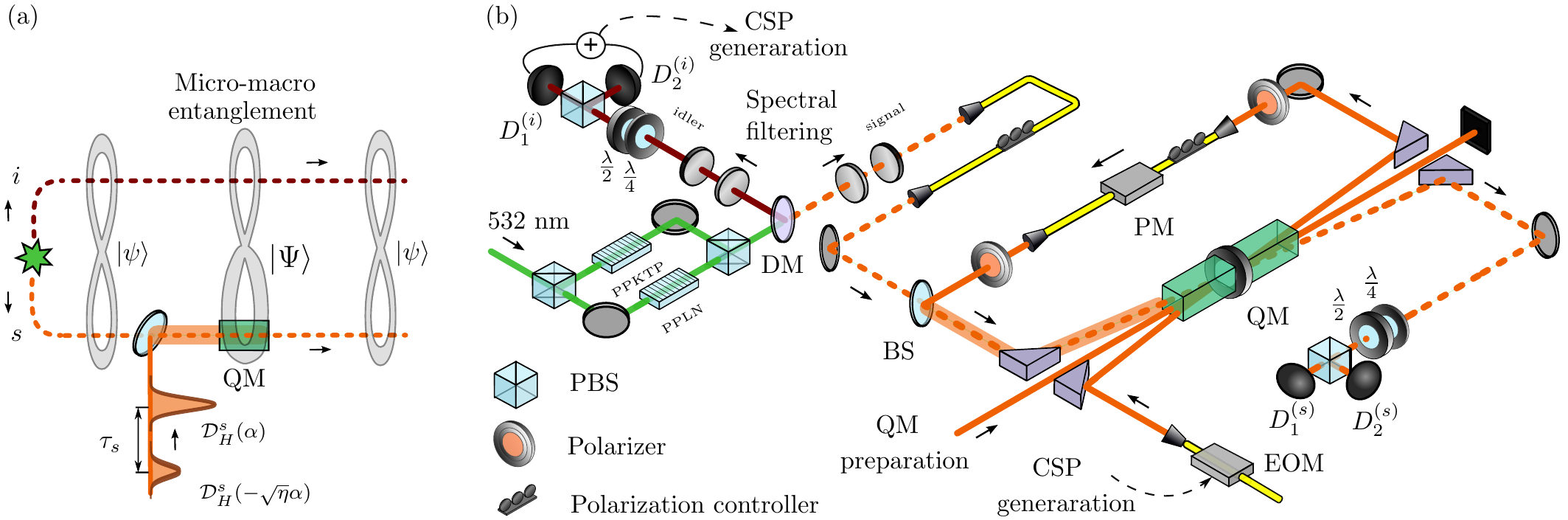}
\caption{Experimental scheme.
(a) Conceptual scheme for the creation and analysis of the light-matter micro-macro entangled state $\ket{\Psi}$. First, a displacement operation $\mathcal{D}^s_H(\alpha)$ is applied on the signal mode of the micro-micro polarization entangled state $\ket{\psi}$ using a beam splitter (BS) with high transmittance. The displaced signal photon of the micro-macro state $\ket{\Psi}$ is then mapped inside a solid-state quantum memory (QM) that has a storage and retrieval efficiency $\eta$. To characterize the state, it is first displaced back to $\ket{\psi}$ (in the ideal case) when it is retrieved from the memory using $\mathcal{D}^s_H(-\sqrt{\eta}\alpha)$, and is then analyzed using various entanglement witnesses.
(b) Detailed setup. A polarization entangled pair of photons is created using spontaneous parametric down-conversion from two periodically poled nonlinear waveguides (PPLN and PPKTP) placed in the arms of a polarization interferometer~\cite{Clausen2014a} seeded by a continuous wave laser (532~nm wavelength). Dichroic mirror (DM) is used to separate two photons spatially. After the spectral filtering the idler photon is detected by one of the detectors ($D_1^{(i)}$ or $D_2^{(i)}$). This event heralds a single photon in the signal mode, and it triggers the generation of a coherent state pulse (CSP) using an electro-optical intensity modulator (EOM) that carves a pulse out of a continuous wave laser at 883~nm. The CSP is sent in the QM in a different spatial mode than the signal mode. This further allows preparing both the displacement and back-displacement pulses with the required delay and amplitudes (see text and SM for details). The relative phase necessary for this is set by an electro-optic phase modulator (PM). The first displacement pulse $\mathcal{D}^s_H(\alpha)$ is synchronized with the heralded single photon on a BS that has a~99.5\% transmittance. 
The resulting state $\ket{\Psi}$ is stored inside the QM and released after a predetermined time of $\tau_s = 50$~ns. The second displacement $\mathcal{D}_H^s(-\sqrt{\eta}\alpha)$ is then applied on the state retrieved from the QM. 
The state is analyzed, together with the idler photon previously measured, using free-space polarization analyzers composed of quarter-wave ($\lambda /4 $) and half-wave ($\lambda /2 $) plates followed by polarizing beam splitters (PBS).  
}
\label{fig:exp}
\end{figure*}

Our experiment is conceptually represented on Fig.~\ref{fig:exp}a. First, an entangled photon pair is generated in the micro-micro state
\begin{equation}
\ket{\psi} = \frac{1}{\sqrt{2}}(\ket{1,0}_s\ket{1,0}_i+\ket{0,1}_s\ket{0,1}_i),
\label{eq:bell1}
\end{equation}
where $s$ and $i$ subscripts are two modes corresponding to the generated \emph{signal} and \emph{idler} single photon, while $\ket{1,0}_{s(i)}\equiv \ket{H}_{s(i)}$ and $\ket{0,1}_{s(i)}\equiv \ket{V}_{s(i)}$ correspond to the horizontal polarization state of the signal (idler) photon and the vertical polarization state, respectively. 
To displace one of the polarization modes of~$s$, the signal photon is superposed with a horizontally-polarized coherent state pulse (CSP) on a highly transmissive beam splitter. This corresponds to a unitary displacement operation $\mathcal{D}^s_{H}(\alpha)$ on the horizontal mode of the signal photon transmitted through the beam splitter~\cite{Paris1996}. The average number of photons contained in the displacement pulse is given by $|\alpha|^2$. After displacement, the entangled state is written as
\begin{equation}
\ket{\Psi} = \frac{1}{\sqrt{2}} \left[ \left(\mathcal{D}^s_H(\alpha) \ket{1,0}\right)_s\ket{1,0}_i + \ket{\alpha,1}_s\ket{0,1}_i\right].
\label{eq:bell2_2}
\end{equation}
This micro-macro entangled state (denoted with a capital $\Psi$ for emphasis) contains a displaced single-photon state of the form $\mathcal{D}(\alpha)\ket{1}$ in the first term, and a coherent state $\ket{\alpha}= \mathcal{D}(\alpha)\ket{0}$ in the second. The idler photon plays the role of the ``micro'' component of the entangled state. Importantly, increasing $|\alpha|$ makes these two terms become more and more distinguishable when using a coarse-grained detector (on the signal mode)~\cite{Sekatski2014a}. This is discussed in detail below.

We use a quantum memory protocol to coherently map the state of the signal mode to the collective state of an ensemble of neodymium atoms frozen in a crystal host~\cite{Afzelius2009a}. This creates $\eta_{\text{abs}}|\alpha|^2$ atomic excitations on average, where $\eta_{\text{abs}}$ is the absorption probability of the quantum memory~(QM). The atomic state obtained after this linear mapping contains the atomic equivalents of the optical states $\ket{\alpha}$ and $\mathcal{D}(\alpha)\ket{1}$~\cite{Frowis2015a}. These atomic states can in principle be directly distinguished using a readout technique that has an intrinsically limited microscopic resolution, as it was shown experimentally in Ref.~\cite{Christensen2014}. Instead, here we analyse the reemission and infer, i.e. indirectly, the atomic state from a model using independent measurements discussed in the text. Thus, after a pre-determined storage time $\tau_s = 50$~ns, the atomic state is mapped back to the optical signal mode. The overall storage and retrieval efficiency is denoted~$\eta$. 
We note that the storage time is much shorter than the {57~\textmu s} coherence time and {300~\textmu s} lifetime of the optical transition. Hence, the collective atomic state is coherent throughout the whole process. 

As part of the measurement of the light-matter entangled state, the state retrieved from the QM is first displaced back with $\mathcal{D}^s_H(-\sqrt{\eta}\alpha)$, where the amplitude is reduced by $\sqrt{\eta}$ to match the limited storage efficiency $\eta$ of the QM.   
To achieve this, an optical pulse is sent through the QM. The timing is such that the part of this pulse that is transmitted (i.e.~not absorbed) by the quantum memory precisely overlaps with the displaced signal photon retrieved from the QM. This is equivalent to overlapping them on a beam splitter that has a limited transmittance, and thus it corresponds to a displacement operation accompanied by loss (see the \suppmat{} for details). In the ideal case, the back-displacement would entirely remove the initial displacement and yield the original micro-micro optical entangled state $\ket{\psi}$. In practice, the displacement back is never perfect in amplitude and phase, which creates noise that limits the maximum size of macroscopic component that can be observed. We note that the displacement happens after the detection of the idler photon, regardless of the measurement outcome. This order however could be reversed by using a mode-locked laser to generate the entangled photons, which would lead to the same results. This independence of time order of the measurements was recently emphasized with a delayed-choice entanglement swapping~\cite{swapp2012}.

We now give more details on the setup; see Fig.~\ref{fig:exp}b. A 532~nm continuous wave laser is coherently pumping two nonlinear waveguides, which probabilistically creates photon pairs at 883~nm (the \emph{signal} photon) and 1338~nm (the \emph{idler} photon). Each photon pair is in superposition of being created in the first waveguide (with horizontal polarizations) and in the second waveguide (with vertical polarizations). Recombination of the output modes of the waveguides leads to a state that is close to the maximally entangled state~(\ref{eq:bell1})~\cite{Clausen2014a}. The spectrum of the idler photon (the signal photon) is filtered to a Lorentzian linewidth FWHM of 240~MHz (600~MHz) using the combination of a Fabry-Perot cavity (etalon) and a highly reflective volume Bragg grating (see Ref.~\cite{Clausen2014a} for details). Due to the strong energy correlation between both photons, the heralded signal photon's (HSP) linewidth is filtered to $\approx 210 $~MHz, corresponding to a coherence time of $\tau_c \approx 1.9$~ns. Detection of the idler photon by detector $D_1^{(i)}$ or $D_2^{(i)}$ heralds a single photon in the signal mode. The detection signal is also used to generate a CSP using an electro-optical intensity modulator which carves a pulse out of a continuous wave laser at 883~nm. 

The quantum memory is based on the Atomic Frequency Comb storage protocol~\cite{Afzelius2009a}. To store light with an arbitrary polarization, we use a configuration consisting of two inline neodymium-doped yttrium orthosilicate crystals \nd{} separated by a half-wave plate. This configuration was previously used to faithfully store polarization qubits~\cite{Clausen2012,Gundogan2012,Zhou2012}, to perform light-to-matter quantum teleportation~\cite{Bussieres2014} and to store hyperentanglement~\cite{Tiranov2015a}. The bandwidth of the prepared quantum memory is 600~MHz and it stores photons for 50~ns with an overall efficiency of~$\eta = 4.6(2)\%$. The back-displacement operation is performed with an interference visibility of 99.85\%, which is remarkably close to being perfect; this is crucial to maximize the size of the displacement.

To quantify how much of the light contained in the displacement pulse is actually displacing the HSP, we must evaluate to what extent their modes are indistinguishable~\cite{Sekatski2014a}. This was done using Hong-Ou-Mandel interference, which was realized as a separate measurement by combining the HSP and the displacement pulse on 50/50 beam splitter (see \suppmat{} for details). A visibility of 74\% was measured and compared to the 85\% expected value. This implies that $0.74/0.85 =87\%$ of the displacement pulse is effectively displacing the single photon~\cite{Sekatski2014a}, and this fraction is used when estimating the number of atomic excitations that are part of the entangled state. 

To reveal the light-matter micro-macro entanglement, we use two methods: the violation of a CHSH-Bell inequality~\cite{Clauser1969} and quantum state tomography.  

\begin{figure}[!t]
	\centering
		\includegraphics[width=0.45\textwidth]{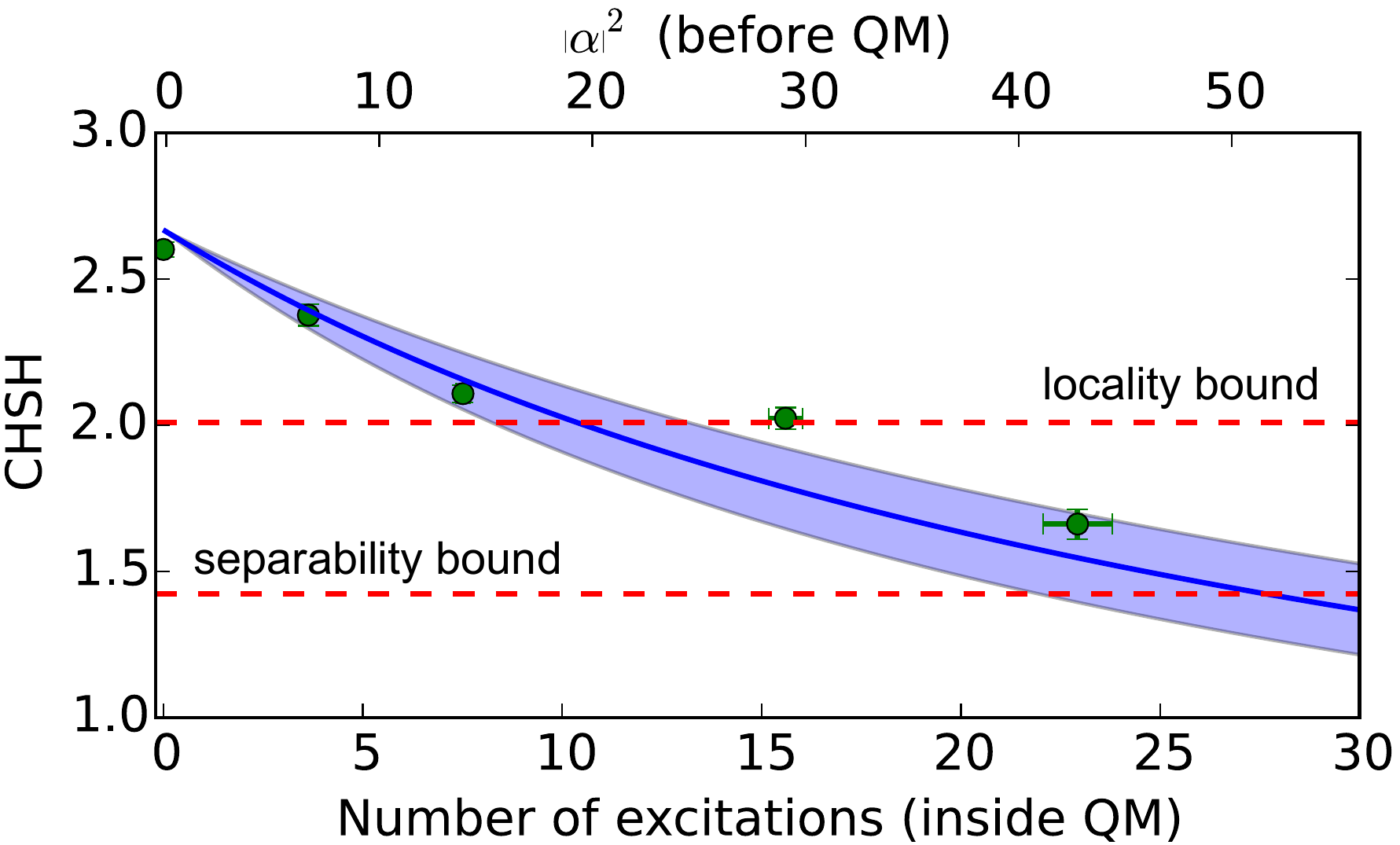}
	\caption{Measured values of the $S$ parameter of the CHSH-Bell inequality (dots) as a function of the size of the displacement before the QM (top $x$-axis) or as a function of the average number of atomic excitations inside the QM (bottom $x$-axis). CHSH violation values are above the local bound with up to $7$ excitations on average, and above the entanglement bound with up to $23$ excitations on average. The error bars are estimated assuming Poisson statistics for the detections. The solid line is obtained from a theoretical model based on independently measured parameters, and the shaded area represents a one standard-deviation uncertainty on the predictions of the model.}
	\label{fig:CHSH}
\end{figure}

We first performed the CHSH test without any displacement operations and obtained a parameter $S = 2.59(3)$, which is above the local bound of~2 by 20 standard deviations. This was then repeated with an increasing displacement size $|\alpha|^2$. The results shown on Fig.~\ref{fig:CHSH} are in a good agreement with a theoretical model based on independently measured experimental parameters (see the \suppmat{} for details). We note that the bases used for all CHSH tests are composed of states of even superposition of $\ket{H}$ and $\ket{V}$. A value of $S=2.099(31)$ is obtained for a displacement containing a mean photon number of $|\alpha|^2 = 13.3(3)$ before mapping the state in the quantum memory. Using the absorption probability $\eta_{\text{abs}} \approx 55\%$, this corresponds to about~7 excited atoms (see the \suppmat{} for details). Interestingly, violating the CHSH inequality shows that the light-matter micro-macro state could lead to strongest form of quantum correlations, namely non-local correlations. Alternatively, the Bell inequality can be used as an entanglement witness if we find $S \ge \sqrt{2} \approx 1.41$~\cite{Roy2005a}. We measured $S=1.65(5)$ with a mean photon number of $\left|\alpha\right|^2 = 42(2)$  before the QM, corresponding to $\approx 23$ excited atoms inside the QM. This is above the separability bound by~4 standard deviations.

\begin{figure}[!t]
	\centering
		\includegraphics[width=0.45\textwidth]{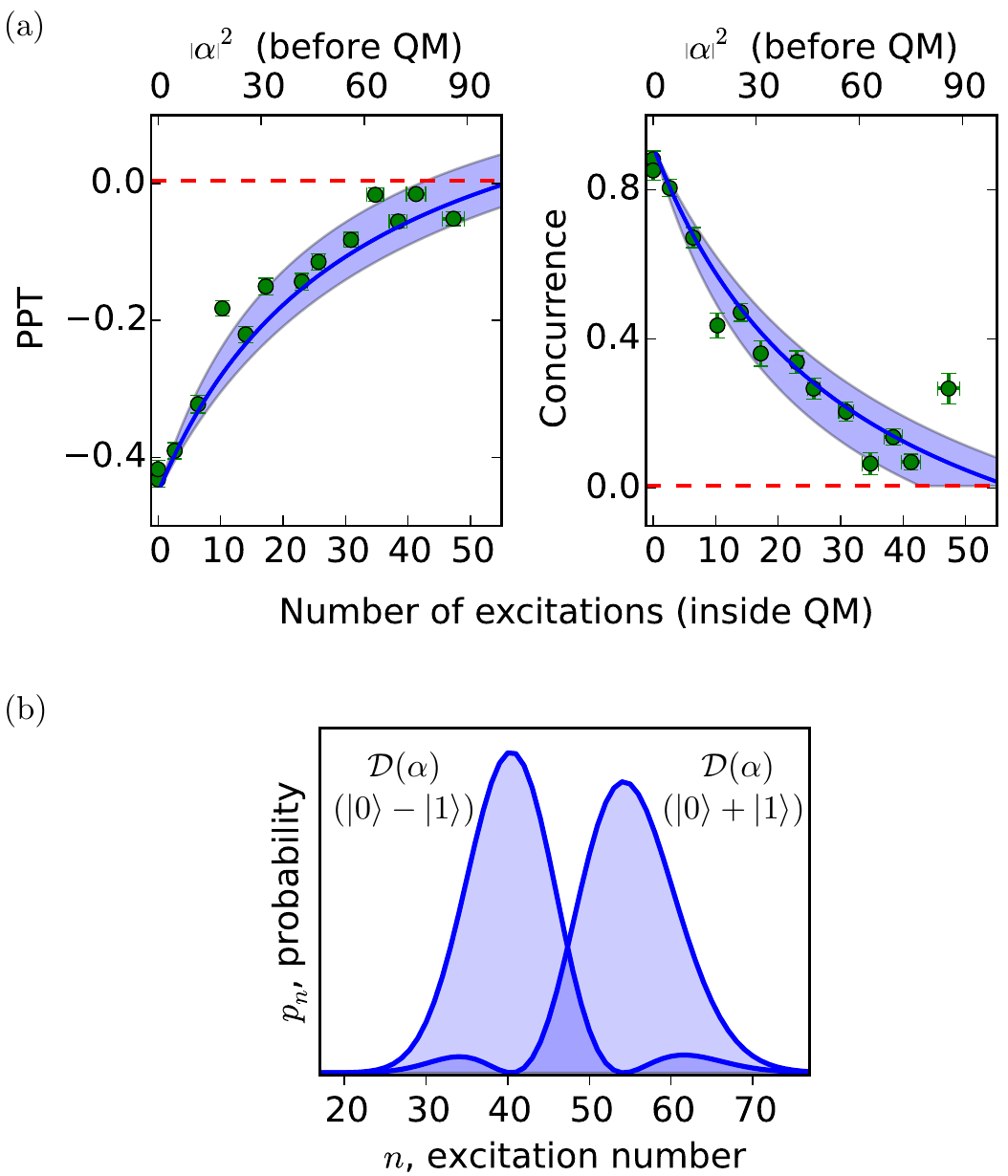}
	\caption{Quantum state characterization. (a) PPT and concurrence values (obtained from quantum state tomography) as a function of the size of the displacement before the QM (top $x$-axis) or as a function of the average number of atomic excitations inside the QM (bottom $x$-axis). The error bars are estimated from Monte-Carlo simulations assuming Poisson noise. The PPT criteria remains negative and the concurrence value remains positive with up to $47$ excitations on average. The solid lines in all graphs are obtained from a theoretical model based on the independently measured parameters, and the shaded areas are the uncertainty on these parameters. (b) Distribution of number of atomic excitations of the two macroscopically distinguishable components $\mathcal{D}(\alpha)\ket{1}$ and $\ket{\alpha}$  when expressed in the $\{\ket{0}+\ket{1},\ket{0}-\ket{1}\}$ basis. }
	\label{fig:tomo}
\end{figure}

To fully characterize the entanglement of the retrieved micro-micro quantum state, we performed an overcomplete set of tomographic measurements and reconstructed the full density matrix.
To prove that the state is still entangled we use two criteria, namely the positivity under partial transposition (PPT)~\cite{Peres1996}, and the concurrence (which is based on the concept of the entanglement of formation)~\cite{Hill1997}. 
Figure~\ref{fig:tomo}a shows results obtained for increasing size of the displacement.
A negative value of $-0.055(10)$ is obtained for the PPT test and a positive concurrence of $0.246(41)$ are obtained for displacements with $|\alpha|^2 = 86(3)$ photons before the QM. This corresponds to~$\approx 47$ excited atoms in the atomic ensemble. These results are in a good agreement with our theoretical model described in the \suppmat{}. 
We attribute the scatter of the data mostly to the fact that it is very sensitive to fluctuations of the visibility of the back-displacement operation. 

The reported light-matter state can be considered as a micro-macro entangled state for the following reason. Let us illustrate first how the size of a given state can be evaluated from the coarse-grained measure presented in Ref.~\cite{Sekatski2014a} by focusing on the state~(\ref{eq:bell2_2}), which can be re-written as
\begin{eqnarray}
&\phantom{-}& \left[\mathcal{D}_H^s(\alpha) (\ket{0}_{sH} + \ket{1}_{sH})\right] (\ket{0}_{sV} \ket{1,0}_i + \ket{1}_{sV}\ket{0,1}_i) \nonumber \\
&-& \left[\mathcal{D}_H^s(\alpha) (\ket{0}_{sH} - \ket{1}_{sH})\right] (\ket{0}_{sV} \ket{1,0}_i - \ket{1}_{sV}\ket{0,1}_i), \nonumber
\end{eqnarray}
where the normalization is omitted. The state therefore involves the superposition of $\mathcal{D}_H^s(\alpha)(\ket{0}+\ket{1})$ and $\mathcal{D}_H^s(\alpha)(\ket{0}-\ket{1})$ in the horizontal mode of the signal photon, and one can obtain one or the other by measuring the idler photon in the basis of diagonal polarizations. 
Although these two components partially overlap in the photon number space, the distance between their mean photon numbers is given by $2|\alpha|$; see Fig.~\ref{fig:tomo}b. For $|\alpha|^2 \gtrsim 2$, they can be distinguished with a single measurement with a probability of $\approx 91$\% using a detector that has a perfect single-photon resolution~\cite{Sekatski2014a}. If measured with a coarse-grained detector, this probability is reduced to 50\% when the coarse graining is of the order of $|\alpha|$ or more. The effective size of the state~(\ref{eq:bell2_2}) can be naturally quantified by the maximum coarse-graining $\sigma_{\max}$ that allows one to distinguish the two components $\mathcal{D}_H^s(\alpha)(\ket{0}+\ket{1})$ and $\mathcal{D}_H^s(\alpha)(\ket{0}-\ket{1})$ with a given probability $P_g$, where $P_g$ should be significantly above 50\% to be meaningful for a single-shot measurement. Similarly, the effective size can be evaluated by comparing the results to an archetypical state involving the superposition of $\ket{0}$ and $\ket{N}$ Fock states, where $N$ is the smallest value that allows distinguishing $\ket{0}$ from $\ket{N}$ with a probability $P_g$ and a coarse graining $\sigma_{\max}$.  
From our results, which are well reproduced by our theoretical model based on independent measurements of the entangled state and experimental parameters, we can confidently give an estimate of the size of the light-matter state from which the entanglement is measured. For $P_g=2/3$, the state is analogous to the state $\ket{\hspace{-2pt}\uparrow}\ket{0} + \ket{\hspace{-2pt}\downarrow}\ket{N}$ with $N \approx 13$, where $\ket{\hspace{-3pt}\uparrow}$ and $\ket{\hspace{-3pt}\downarrow}$ represent microscopic orthonormal states.

Naturally, one must also carefully consider the effect of loss in the signal mode before the beam splitter used for the displacement, as well as the absorption probability in the QM. In the \suppmat{} we show that if the heralding probability to find the signal photon at the beam splitter is $\eta_h$ and the absorption in the QM is $\eta_{\text{abs}}$, the displacement creates a mixture of the state $\ket{\Psi}$ with a displacement of amplitude $\sqrt{\eta_{\text{abs}}}\alpha$ with probability $\eta_h\eta_{\text{abs}}$, and a separable state with the complimentary probability. In our case we have $\eta_h\eta_{\text{abs}} \approx 10\%$, which makes the two macroscopic states nearly indistinguishable, even with a detector with perfect microscopic resolution. This exemplifies that the direct observation of macroscopic features is a very challenging task. Nevertheless, we stress that the entanglement signature that we directly observe is stemming from the micro-macro entangled state component of the mixture, whose effective size is defined as above. The observation of this entanglement, and its behaviour with increasing size, is the main result of this Letter. The direct observation of the size of the superposition with an actual coarse-grained detector is left for future work. This would require reduced loss and a highly efficient quantum memory. Achieving this is certainly conceivable, given the large storage efficiencies that can now be obtained with some quantum memories~\cite{Bussieres2013} and with the progress of linear detectors to achieve sub shot-noise resolution (e.g.~see~\cite{Gerrits2012a}). Homodyne detection could also prove useful for distinguishing the states, as demonstrated in Ref.~\cite{Lvovsky2013a}. Overall, our approach could certainly be improved with other types of quantum memory, which has the potential to yield larger quantum superpositions in matter.

\section*{Acknowledgements}

We thank Florian Fr\"owis, Jean Etesse, Anthony Martin, Natalia Bruno and Pavel Sekatski for useful discussions, as well as Claudio Barreiro and Rapha\"el Houlmann for technical support. We thank Harald Herrmann and Christine Silberhorn for lending the PPLN waveguide.

\section*{Funding Information}

This work was financially supported by the European Research Council (ERC-AG MEC) and the  National Swiss Science Foundation (SNSF) (including grant number PP00P2-150579). J.~L.~was supported by the Natural Sciences and Engineering Research Council of Canada (NSERC).

\bibliography{displacement}

\newpage

\newpage

\appendix
\section*{Appendix for ``Light-matter micro-macro entanglement''}

\section*{Overview}

In this \suppmat,  we provide details on our experiment and we describe the theoretical model.   Section~\ref{sec:displ}  describes
the details of the implementation of the displacement operations.  Section~\ref{sec:hom} presents the Hong-Ou-Mandel dip experiment to prove the indistinguishability of the heralded single photon and the coherent state pulse. Section~\ref{sec:theory} presents two theoretical models to compare to our experimental results.

\section{Implementation of the displacement operation} 
\label{sec:displ}

\subsection*{Displacement with an arbitrary polarization state}
Here we show that the polarization state of the displacement pulse can be arbitrary without changing the description of our experiment. Let us consider the case where the polarization of the displacement is expressed as $\ket{\psi} = \alpha \ket{H} + \beta \ket{V}$. The displacement is applied to $\frac{1}{\sqrt{2}}
\left(\left|H\right\rangle_s\left|H\right\rangle_i + e^{i\theta}\left|V\right\rangle_s\left|V\right\rangle_i\right)$. We note that
\begin{eqnarray}
&\phantom{=}&\frac{1}{\sqrt{2}}\left(\left|H\right\rangle_s\left|H\right\rangle_i + e^{i\theta}\left|V\right\rangle_s\left|V\right\rangle_i\right) \nonumber \\
&=&\frac{1}{\sqrt{2}}\left(\left|\psi\right\rangle_s\left|\phi\right\rangle_i + e^{i\theta}\left|\psi^{\perp}\right\rangle_s\left|\phi^{\perp}\right\rangle_i\right),
\label{eq:bell_arb_sm}
\end{eqnarray}
where $\left|\phi\right\rangle = \alpha^{*}\left|H\right\rangle + e^{i\theta}\beta^{*}\left|V\right\rangle$, $\braket{\phi}{\phi^{\perp}} = \braket{\psi}{\psi^{\perp}}=0$. We see directly that a displacement with polarization $\ket{\psi}$ will produce a state equivalent to displacing $\frac{1}{\sqrt{2}} \left(\left|H\right\rangle_s\left|H\right\rangle_i + e^{i\theta}\left|V\right\rangle_s\left|V\right\rangle_i\right)$ with a horizontal polarization.

\subsection*{Displacement with a quantum memory}
\begin{figure}[!t]
\centering
\def\svgwidth{0.35\textwidth}
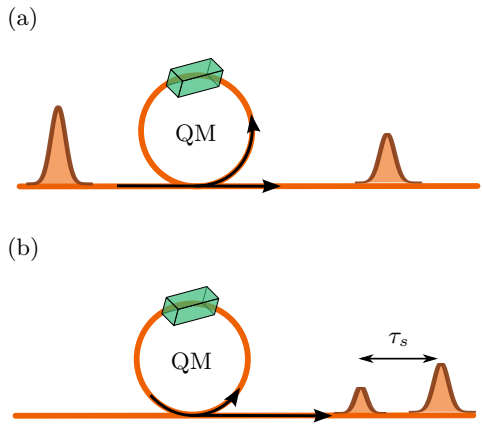
\caption{Quantum memory represented as a storage loop. For our purpose, the AFC quantum memory can be seen as a light-matter beam splitter that loops back onto itself, and the length of the loop corresponds to the storage time $\tau_s$ of the QM. (a) When an optical pulse is incident on the QM, it acts like a beam splitter that splits the light on a optically transmitted part and onto the collective state of the atomic ensemble. (b) At the time of re-emission, the QM acts again like a beam splitter that maps the collective atomic state to the optical mode and back onto itself. The number of photons re-emitted is thus accompanied by the corresponding reduction in the number of atomic excitations in the QM. 
}
\label{fig:qm}
\end{figure}

Here we explain how the quantum memory (QM) can be used to realize the displacement operations. For our purpose, the QM can be seen as a storage loop. As shown on Fig.~\ref{fig:qm}a, light incident on the QM is either absorbed with probability $\eta_{\text{abs}} \approx 55\%$ or transmitted with probability $1-\eta_{\text{abs}}$. This is physically equivalent to a beam splitter whose outputs are the directly transmitted optical mode and the atomic mode onto which light is linearly mapped to a collective atomic state~\cite{Afzelius2009a}. 

The AFC storage protocol that we use is such that the re-emission process occurs when all the spectral components of the collective atomic state are in phase, which in our case happens at $\tau_s = 50$~ns after absorption~\cite{Afzelius2009a}. However, this rephasing is not perfect, and therefore only part of the atomic excitations are converted back to light. In the storage loop representation, the atomic mode is looped back onto the ``light-matter'' beam splitter after the storage time (see Fig.~\ref{fig:qm}b), albeit with a beam splitting ratio that is different than the one of the absorption process, which is due to the details leading to the rephasing. 

\begin{figure*}[!t]
\centering
\def\svgwidth{1.00\textwidth}
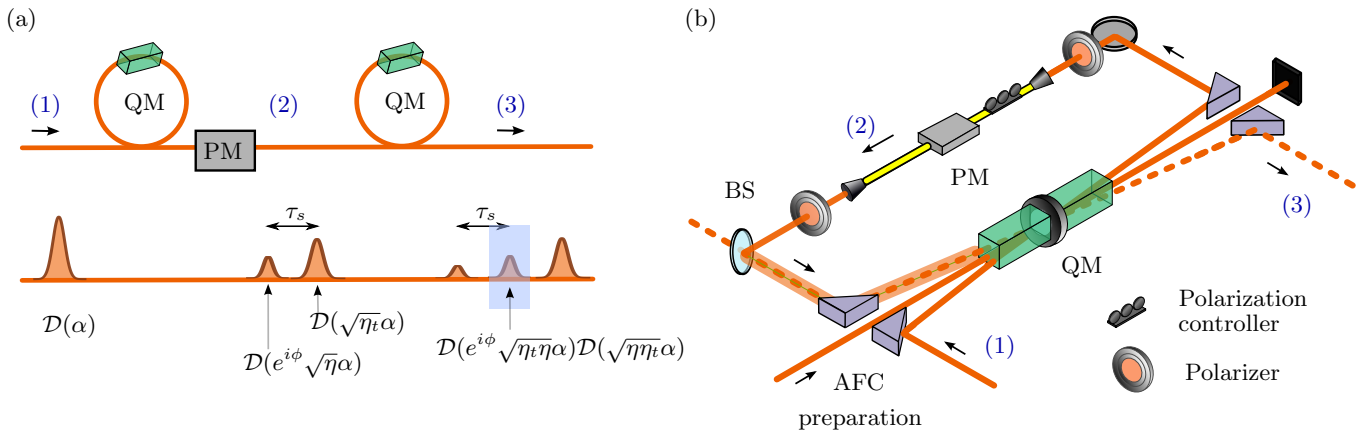
\caption{
(a) presents the conceptual evolution of the pulses used for the displacements, and (b) shows how it was implemented. (1) A coherent state pulse is generated by the detection of the idler photon. It corresponds to the displacement $\mathcal{D}(\alpha)$ applied on vacuum. This pulse is incident on the QM in an optical mode that is distinct from the one used to store the signal mode. (2) After the QM and the phase modulator (PM), we obtain two pulses corresponding to the displacements $\mathcal{D}(\sqrt{\eta_t}\alpha)$ and $\mathcal{D}(\e^{i\phi}\sqrt{\eta}\alpha)$, where $\eta_t$ and $\eta$ are the transmission probability and storage efficiency of the QM, respectively, and $\phi$ is the phase applied on the second pulse by the PM. (3) After the second passage through the QM, we have three pulses. The middle one corresponds to the displacement $\mathcal{D}(\e^{i\phi}\sqrt{\eta_t \eta}\alpha)\mathcal{D}(\sqrt{\eta \eta_t}\alpha)$, and is the one that is applied on the displaced signal photon retrieved from the QM. We see that the amplitudes are perfectly balanced, and setting $\phi = \pi$ will yield destructive interference, which constitutes the back displacement.
}
\label{fig:exp_sm}
\end{figure*}

To realize a near-perfect back displacement operation one has to achieve the lowest phase and amplitude noise between the two displacements. This is especially difficult to realize with free-space optics for long delays such as 50~ns. However, we note that 
the QM itself can be used to create the two pulses; see Fig.~\ref{fig:exp_sm}. First, the coherent state pulse (CSP), created whenever an idler photon is detected, is sent through the QM in a spatial mode that is distinct from the one of the signal photon. This creates two pulses delayed by the storage time of the QM ($\tau_s=50$~ns). The pulses then go through a fibre-pigtailed electrooptic phase modulator (PM) applying a phase of $\phi = \pi$ on the second one. The first displacement operation is performed by combining the heralded signal photon and the first component of the optical signal emerging from the PM. For this we use a non polarizing beam splitter with a transmittance $T = 99.5$\%, resulting in a lossless displacement operation. 
The back displacement operation corresponds to the interference between the part of the second pulse emerging from the PM that is directly transmitted by the QM, and the part of the atomic excitation that is converted into the optical mode. Because the probability to map the atomic state into the optical state is smaller than 100\%, this corresponds to a displacement operation with some loss into the atomic mode. 
To quantify the quality of the back displacement, we measured the visibility of the interference between the two displacements with the signal photon blocked. We obtained an average visibility of 99.85(2)\%, which is very close to being perfect. 

To create a representative theoretical model of our experiment, we characterized the polarization of the light that is remaining after back displacement operation. For this the photon pair source was blocked and polarization state tomography was performed on the weak coherent state. We expected to find a pure polarization state, because the remaining light is expected to comes from the well polarized CSP. Surprisingly, we found that the polarization state is almost completely depolarized, having a fidelity of 97(1)\% with completely depolarized state. This indicates that the limit to the visibility of the back displacement is at least partly due to some process that we could not identify. This will require further investigation.

\subsection*{Effect of loss}
We now describe the effect of loss in our experiment. The first loss to consider is the heralding efficiency $\eta_h$, which is the probability to find the heralded signal photon at the beam splitter used for the displacement. If the photon is lost, the displacement pulse is applied on vacuum $\ket{0}$ rather than on $\ket{1}$. 
In this case, the state $\rho$ created is a mixture of the desired micro-macro entangled state $\ket{\Psi} = \mathcal{D}_H^s(\alpha)\ket{\psi}$, where
\begin{eqnarray}
\ket{\psi} &=& \frac{1}{\sqrt{2}}(\ket{1,0}_s\ket{1,0}_i+\ket{0,1}_s\ket{0,1}_i), \nonumber \\
\ket{\Psi} &=& \frac{1}{\sqrt{2}} \left[ \left(\mathcal{D}^s_H(\alpha) \ket{1,0}\right)_s\ket{1,0}_i + \ket{\alpha,1}_s\ket{0,1}_i\right], \nonumber 
\label{eq:bell1a}
\end{eqnarray}
with a separable component: 
\begin{eqnarray}
\rho = \eta_h \ket{\Psi}\bra{\Psi} &+& (1-\eta_h) \ket{\alpha, 0}_s\bra{\alpha, 0} \otimes \frac{\mathbb{I}_i}{2} \label{eq:macro1}
\end{eqnarray}
where $\mathbb{I}_i$ is the $2\times 2$ identity matrix representing a completely mixed polarization for the idler photon. 

The loss caused by the finite absorption probability in the QM then reduces the size of the displacement and create another separable component to the mixture. Using $\eta_{\text{abs}}$ and $\eta_t = 1-\eta_{\text{abs}}$ to denote the absorption and transmission probabilities, one can show that the light-matter entangled state obtained is
\begin{eqnarray}
\rho' &=& \eta_{\text{abs}} \eta_h \mathcal{D}_H^s(\sqrt{\eta_{\text{abs}}}\alpha)\ket{\psi}\bra{\psi}{\mathcal{D}_H^s}^{\dagger}(\sqrt{\eta_{\text{abs}}}\alpha) \nonumber \\
&+& (1-\eta_h\eta_{\text{abs}}) \ket{\sqrt{\eta_{\text{abs}}}\alpha, 0}_s\bra{\sqrt{\eta_{\text{abs}}}\alpha, 0} \otimes \frac{\mathbb{I}_i}{2}.
\label{eq:werner_vac}
\end{eqnarray}
The first term corresponds to the desired light-matter micro-macro entangled state, where the amplitude of the displacement is given by $\eta_{\text{abs}}|\alpha|^2$. The other term is separable, and do not contribute to detected signature of  entanglement. Rather, they create noise that masks the entanglement signature, which reduces the maximum size of the displacement that we can apply.  

\begin{figure}[!t]
\centering
\def\svgwidth{0.42\textwidth}
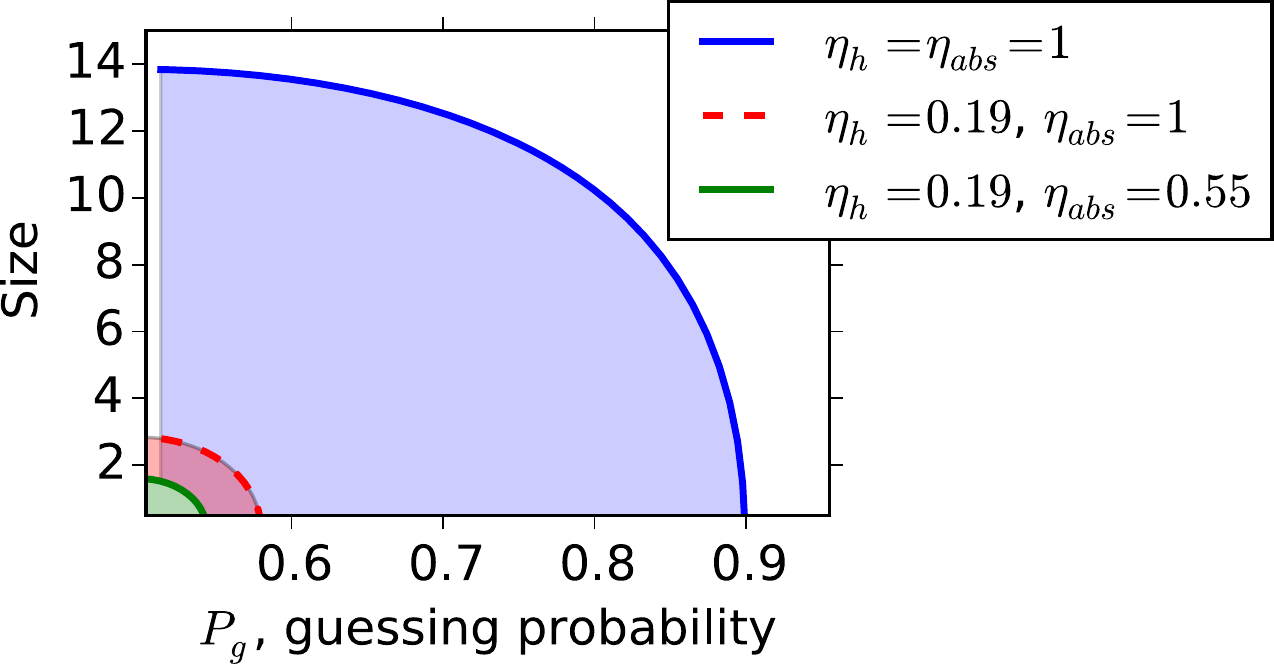
\caption{
Size of the light-matter entangled states $\rho'$ (Eq.~\ref{eq:werner_vac}) 
with $|\alpha|^2 = 47$ atomic excitations on average as a function of the guessing probability $P_g$ to distinguish two macroscopic components of state $\rho'$ for different values of $\eta_h$ and  $\eta_{\text{abs}}$.
}
\label{fig:size_qm_sm}
\end{figure}

One can consider how the loss would affect our ability to directly observe the distinguishability of the macroscopic states of the superposition using a coarse-grained detector. 
On Fig.~\ref{fig:size_qm_sm} we show how loss affects the effective size of the superposition in the cases $\eta_h = 1$ (no loss), $\eta_h = 0.19$ (before the QM) and $\eta_h = 0.19$ with $\eta_{\text{abs}} = 0.55$ (inside the QM). The loss and finite absorption probabilities reduce the maximum probability to distinguish the two macroscopic states to a value that is $\approx 53\%$ with a detector that has a perfect single-photon resolution. This exemplifies that the direct observation of the distinguishability of the macroscopic components requires maximizing $\eta_h$ and $\eta_{abs}$, which is a challenging but conceivable task for future work.

\subsection*{Double detection events}
We note that in all of our measurements, the probability to detect two photons in the signal mode was at least fifty times smaller than detecting a single photon (which was obtained for the case of 47 atomic excitations). Hence, double detection events could essentially be ignored.

\section{Hong-Ou-Mandel interference}
\label{sec:hom}

\begin{figure}[!t]
\centering
\def\svgwidth{0.5\textwidth}
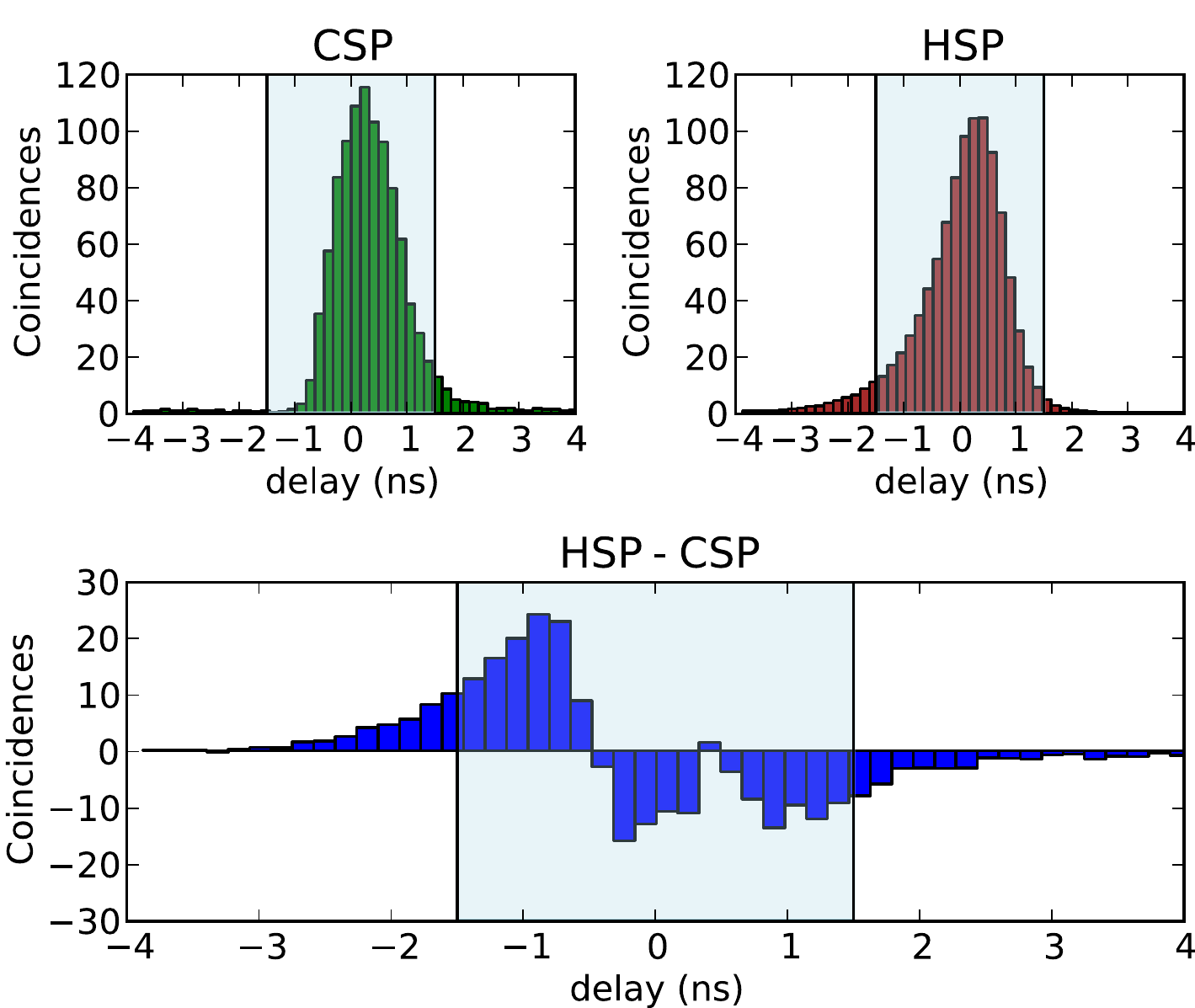
\caption{Temporal profiles of the CSP and HSP. (a) Temporal mode of the coherent state pulse (CSP) shaped from CW laser using electro-optical modulator (EOM).
(b) Temporal profile of the heralded single photon (HSP). (c) Difference between two histograms corresponding to the CSP and HSP, where each histogram was normalized. A coincidence window of 3~ns (shaded regions) was used for further analysis.
}
\label{fig:overlap}
\end{figure}

\begin{figure}[!t]
\centering
\def\svgwidth{0.5\textwidth}
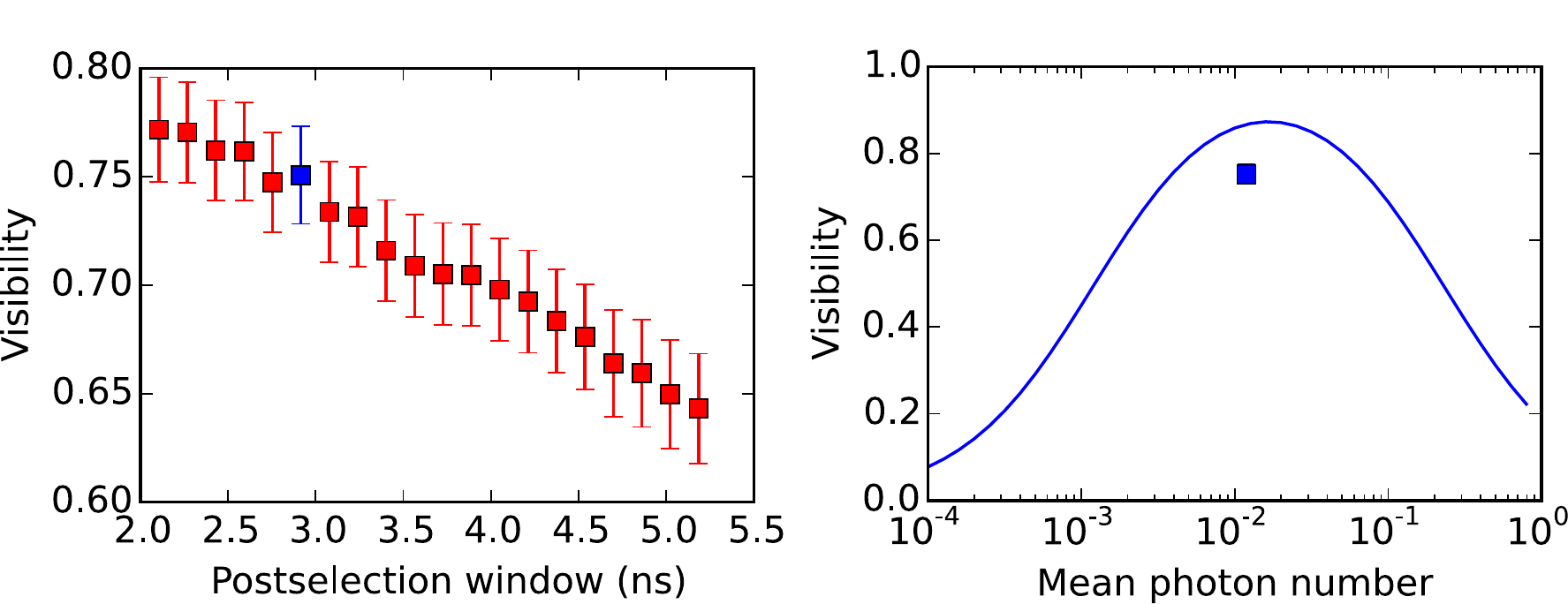
\caption{Hong-Ou-Mandel dip measurement between HSP and CSP. 
(a) Measured visibility $V_{\text{m}}$ of the HOM dip as a function of coincidence window. A window of 3~ns was used during the experiment. The visibility goes down as one increases the postselection window due to the lower overlap between the temporal modes of HSP and CSP.
(b) Comparison between experimentally measured visibility of HOM dip (square) and theoretical model (solid line). The mean photon number of the CSP was 0.012(1) and the creation probability of the photon pair was 0.005, while the heralding efficiency was equal to 19\%. Since the photon number statistics of HSP and CSP are different the maximum visibility of HOM dip between them for the given photon pair creation probability can be obtained only for certain mean photon number $\mu$ of the CSP. The visibility for the low $\mu$ of the CSP is limited by the multi photon creation from the photon pair source while for the high $\mu$ the higher terms of Poisson distribution of the CSP will reduce the maximum visibility~\cite{Bussieres2014}.
}
\label{fig:hom}
\end{figure}

To probe for the indistinguishability between the heralded single photon (HSP) and the coherent state pulse (CSP), a Hong-Ou-Mandel (HOM) type interference experiment was performed. This characterization is important to determine the size of the displacement. The reason is simple: if the displacements are not applied on the mode of the HSP, then the single photon is not displaced at all. The case where the modes of the CSP and HSP overlap only partially was theoretically considered in Ref.~\cite{Sekatski2012b}, where it is shown that the ratio $R=V_{\text{m}}/V_{\text{e}}$ between measured $V_{\text{m}}$ and expected $V_{\text{e}}$ HOM visibilities gives a lower bound on the fraction of the CSP that actually displaces the HSP. Hence, if the size of the CSP is $|\alpha|^2$, then the size of the displacement is $R|\alpha|^2$.

In our case, the main mismatch between the modes of the CSP and HSP is due to their temporal modes; see Fig.~\ref{fig:overlap}. The mode of the HSP is determined by the energy correlation between the signal and idler photons, combined with filtering bandwidths used in the source of entanglement; see details in Ref.~\cite{Clausen2014a}. 
The CSP has an almost Gaussian temporal profile, which is defined by the high speed pulse generator used to drive the electrooptic EOM that is carving the pulse out of a continuous wave laser at 883~nm. For comparison, the temporal modes of the CSP and the HSP are shown in Fig.~\ref{fig:overlap}. 

To measure the HOM dip visibility, the HSP and CPS are combined on a 50/50 beam splitter and synchronized in time. To extract the visibility of the HOM dip, we measure the coincidence rate with the polarization of the CPS either parallel $R_{\|}$ or perpendicular $R_{\perp}$ to the HSP. The measured visibility is given by $V_{\text{m}} = (R_{\perp} - R_{\|})/R_{\perp}$. This value depends on the temporal width of the coincidence window. This is because using a short window compared to the width of the temporal modes erases differences between them, which increases their indistinguishability, but it also reduces $\eta_h$ and the detection rate. The value of the measured visibility as a function of the width of the coincidence window is shown in Fig.~\ref{fig:hom}a. A tradeoff value was chosen with a coincidence window of 3~ns, which yielded an heralding efficiency of $\eta_h = 0.19$.

The expected visibility $V_{\text{e}}$ is calculated taking into account the heralding efficiency, the photon pair creation probability $p = 5\times 10^{-3}$ inside the coincidence time window of 3~ns, as a function of the mean photon number contained in the CSP.
The result of this calculation is shown on Fig.~\ref{fig:hom}b. For the value $|\alpha|^2 = 0.012$ used with the 3~ns window on Fig.~\ref{fig:hom}a, the expected visibility is $V_{\text{e}} = 85\%$. The measured visibility is $V_{\text{m}}= 74(2)\%$, which shows that we do not have perfect overlap between the HSP and CSP. The difference is due partly to the imperfect spectral-temporal modes overlap between two waveforms, and also to the fluctuations on the central frequency of the HSP, which is due to the small fluctuations on frequency of the pump laser at 532~nm (the latter are caused by the frequency stabilization mechanism that we need to apply to get frequency correlated photons, see~\cite{Clausen2014a}). 

The measured ratio $R = 0.74/0.85 = 87\%$ is used to correct the size of the displacements that are presented in this work.

\section{Theoretical model of the experiment}
\label{sec:theory}
\subsection*{Simple theoretical model}

Here we derive a simple theoretical model to compare to our experimental results. First, we include in our description that actual micro-micro entangled state $\rho_{mm}$ (obtained without any displacement) is itself not an ideal pure state, but can rather be approximated by the Werner state
\begin{equation}
\rho_{mm} = V_{mm} \ket{\psi}\bra{\psi} +  (1-V_{mm})\frac{\mathbb{I}}{4}
\end{equation}
where $\mathbb{I}$ is the $4\times4$ identity matrix, and $V_{mm} \approx 94\%$ is the entanglement visibility of the micro-micro entangled state. Next, to include the contribution of the displacement in our description, we recall that we found that light remaining after the imperfect back displacement was almost entirely depolarized. Hence, the actual state $\rho_{mm}'$ emerging from the QM can be seen, at first approximation, as a mixture of $\rho_{mm}$ (obtained when the signal photon was not lost) and the completely mixed two-qubit state obtained when the signal photon is replaced with a completely mixed polarization photon. We have
\begin{eqnarray}
\rho_{mm}' &=& (1-\epsilon) \rho_{mm} + \epsilon \frac{\mathbb{I}}{4} \nonumber \\
&=& V_{mm}(1-\epsilon) \ket{\psi}\bra{\psi} + [1-V_{mm}(1-\epsilon)]\frac{\mathbb{I}}{4} \label{eq:werner_si}
\end{eqnarray}
where we denote $\epsilon$ as the noise probability. 

We then need to properly model $\epsilon$ as a function of independently measured experimental parameters. Assuming the interference visibility $V$ between two displacements operations, one can calculate $\epsilon$ assuming a Poisson photon number statistics. 
On the one hand, the probability to detect a photon from the noisy background is proportional to 
\begin{equation}
p_n = \e^{-\mu} \sum_{n=1}^{\infty} \frac{\mu^n}{n!}\left[1- \left(1- 2 \eta  (1-V) \right)^n \right],
\label{eq:model1}
\end{equation}
where $\eta$ is the efficiency of the quantum memory and $\mu$ is the mean photon number in the back displacement operation. 
On the other hand, the probability that no photon (other than the heralded signal photon) leaks out of the QM after the back displacement is   
\begin{equation}
\bar{p}_{n} = \e^{-\mu} \sum_{n=1}^{\infty} \frac{\mu^n}{n!} \left[1- 2 \eta (1-V) \right]^n.
\label{eq:model2}
\end{equation}
Finally, the probability to detect the heralded signal photon while no photon from the noisy background leaks out is proportional to
 \begin{equation}
p_{s} = \eta_h T \eta\, \bar{p}_n,
\label{eq:model3}
\end{equation} 
where $\eta_h = 0.19$ is the heralding efficiency of the signal photon and $T = 99.5\%$ is the transmission of the first beam-splitter. Using these definitions, the noise probability $\epsilon$ of the Werner state~(\ref{eq:werner_si}) can be expressed as 
\begin{equation}
\epsilon = \frac{p_n}{p_{s}+p_n}.
\label{eq:model4}
\end{equation} 
Using the independently measured parameters $\eta_h=19(2)\%$, $T=99.5\%$, $\eta=4.6(2)\%$, $V = 99.85(2)\%$ and $V_{mm} = 94\%$, we can estimate the expected $S$-parameter value for the CHSH violation as a function of the size of the displacement. We can also predict the values for the PPT criterion and the value of the concurrence (Fig.~2 and 3 of the main text).

\subsection*{Detailed theoretical model}

Here we derive a more detailed model for predicting the expected value of the CHSH $S$ parameter, which can be compared with the result of the simple model. We show that both models produce essentially the same results, and they both correspond very well with our experimental results. This strengthens our claim. 

This model starts with a description of the spontaneous parametric down conversion source which produces photons in coupled modes, labelled by the bosonic operators $a$ and $b$. 
The photons are created in maximally entangled states in polarization, meaning that each mode splits into two orthogonal polarizations $a-a_{\perp}$ and $b-b_{\perp}$. The Hamiltonian of such a process is $\mathcal{H} = i \chi (a^{\dagger}b^{\dagger}_\perp -a^{\dagger}_\perp b^{\dagger} + \text{h.c})$, where $\chi$ is proportional to the
non-linear susceptibility of the crystal and to the power of the pump. The expression of the corresponding state $\left|\psi \right\rangle$ is obtained by applying $e^{-i\mathcal{H} t}$ on the vacuum $\left|\underline{0} \right\rangle$ , as
we are focusing on spontaneous emissions (0 is underlined to indicate that all modes are in the vacuum). It can be written as~\cite{Vivoli2015a}
\begin{equation}
\left|\psi\right\rangle = \left( 1-T_g^2 \right) e^{T_g a^\dagger b^{\dagger}_{\perp}} e^{T_g a^{\dagger}_{\perp} b^{\dagger}} \left|\underline{0} \right\rangle,
\label{eq:app_1}
\end{equation}
where $T_g =\tanh g$, $g = \chi t$ being the squeezing parameter.

For the detectors, we used non-photon number resolving detectors with non-unit efficiency $\eta$ and dark count
probability $p_{\text{dc}}$. The no-click event for the mode $a$ for  example, is associated to a positive operator~\cite{Vivoli2015a}
\begin{equation}
D^a_{\text{nc}} = (1-p_{\text{dc}})(1-\eta)^{a^\dagger a},
\label{eq:app_2}
\end{equation}
while the click event corresponds to $D^a_\text{c} = \mathbbm{1} - D^a_{\text{nc}}$.

We now calculate the conditional state in modes $b$, $b_\perp$ once the modes $a$, $a_{\perp}$ are detected. Note first that the structure of the Hamiltonian is such that we can consider that for any measurement choice, $a$ and $a\perp$ are the
eigenmodes of the measurement device. The state that is conditioned on a click in $a$ and no click in $a_{\perp}$ is given by 
\begin{multline}
\rho_{b,b_\perp} = (1 - p_{\text{dc}}) \text{tr}_{a,a_\perp} \left((1 - \eta)^{a^\dagger a} \otimes \mathbb{1}_{a_\perp} \left| \psi \right\rangle \left\langle \psi \right| \right) - 
\\ - (1 - p_{\text{dc}})^2 \text{tr}_{a,a_\perp} \left((1 - \eta)^{a^\dagger a} \otimes (1-\eta)^{a^\dagger_\perp a^\dagger} \left| \psi \right\rangle \left\langle \psi \right| \right).
\label{eq:app_3}
\end{multline}

Using $x^{a^\dagger a} f(a^\dagger) = f(xa^\dagger)x^{a^\dagger a}$, we get~\cite{Vivoli2015b}
\begin{multline}
\rho_{b,b_\perp} = (1 - p_\text{dc}) \frac{1 - T_g^2}{1 - T_g^2R^2} \rho^{b}_{\text{th}}(Tg) \rho^{b_\perp}_{th}(RT_g) - 
\\ -(1 - p_\text{dc})^2 \frac{\left(1 - T_g^2 \right)^2}{\left(1 - T_g^2R^2\right)} \rho^{b}_{\text{th}}(RT_g) \rho^{b_\perp}_{th}(RT_g)
\label{eq:app_4}
\end{multline}
with $\rho^b_{\text{th}}(T_g) = \left(1 - T_g^2\right) \sum_{n=0}^{+\infty} T_g^{2k} \left| k \right\rangle_b \left\langle k \right|$, $\rho^{b_\perp}_{\text{th}}(RT_g) = \left(1 - (RT_g)^2\right) \sum_{n=0}^{+\infty} (RT_g)^{2k} \left| k \right\rangle_{b_\perp} \left\langle k \right|$ ... i.e. the conditional state can be written as a difference between products of thermal states.

As thermal states are classical states, they can be written as a mixture of coherence states, that is 
\begin{equation}
\rho^b_\text{th}(T_g) = \int{ d^2 \gamma P^{\bar{n}} (\gamma) \left| \gamma \right\rangle_{b} \left\langle \gamma \right|}
\label{eq:app_5}
\end{equation}
with $P^{\bar{n}}(\gamma)  = \frac{1}{\pi \bar{n}} e^{-\left| \gamma \right|^2 / \bar{n}} $ and $\bar{n} = T_g^2/(1 - T_g^2)$. The result of Bob measurements can thus be deduced from the distribution of results obtained with coherent states. In particular, a state
\begin{equation}
\left| \alpha \right\rangle_b \left\langle \alpha \right| \otimes \left| \beta \right\rangle_{b_\perp} \left\langle \beta \right|
\label{eq:app_6} 
\end{equation}
becomes
\begin{equation}
\left| \bar{\alpha}+\cos\theta M \right\rangle_b \langle \underbrace{\bar{\alpha}+\cos\theta M}_{\hat{\alpha}} |  \otimes  | \underbrace{\bar{\beta}+\cos\theta M}_{\hat{\beta}} \rangle_{b_\perp}  \left\langle \bar{\beta}+\cos\theta M \right|
\label{eq:app_7} 
\end{equation}
after the memory where $\bar{\alpha} = \alpha T_1 T_2 \sqrt{\eta_c}\sqrt{\eta_d}$ (similarly for $\bar{\beta}$) accounts for the transmission of the beamsplitter used for the displacement operation, $\eta_c$ for the coupling efficiency and... 
$M = iT_2\gamma \phi$ where $\gamma^2$ is the size of the displacement and $\phi$ is the error on the relative phase between the two displacements. It is averaged out with gaussian noise to account for the limited accuracy of our back displacement operation. $\theta$ accounts for the angle between the measurement setting of Alice and the polarization of the laser used to implement the displacement operation. For Bob's measurement setting with eigenmodes $b^{\theta{'}} = \cos \theta{'}b + \sin \theta{'} b_{\perp}$ and  $b^{\theta{'}}_\perp = \sin \theta{'}b - \cos \theta{'} b_{\perp}$,
the probability to get no click in $b^{\theta{'}}$ and one click in $b^{\theta{'}}_\perp$
with the state (\ref{eq:app_7}) is given by
\begin{multline}
\text{tr}_{b^{\theta{'}}, b^{\theta{'}}_\perp} \left[ (1 - \eta_d)^{b^{\theta{'}\dagger}, b^{\theta{'}}} \otimes 
( 1 - (1 - \eta_d)^{b^{\theta{'}\dagger}, b^{\theta{'}}} ) \right.
\\ \left. \left| \cos \theta{'} \hat{\alpha} + \sin \theta{'} \hat{\beta} \right\rangle_{b{'}} \left\langle \cos \theta{'} \hat{\alpha} + \sin \theta{'} \hat{\beta} \right| \right.
\\ \left. \left| \sin \theta{'} \hat{\alpha} - \cos \theta{'} \hat{\beta} \right\rangle_{b{'}_\perp} \left\langle \sin \theta{'} \hat{\alpha}  \cos \theta{'} \hat{\beta} \right|  \right]
\\ = e^{ - \left| \cos \theta{'} \hat{\alpha} + \sin \theta{'} \hat{\beta}\right|^2\eta_d} (1 - e^{ - \left| \sin \theta{'} \hat{\alpha} - \cos \theta{'} \hat{\beta}\right|^2\eta_d}).
\label{eq:app_8}
\end{multline}
Attributing the result $+1$ to the events \{no click in a and one click in $a_\perp$\} and \{no click in $b^{\theta{'}}$ and one click in $b^{\theta{'}}_\perp$\} the joint probability $p(+1 + 1| \theta \theta{'})$ can be obtained the previous result together with Eqs.~(\ref{eq:app_4}) and (\ref{eq:app_5}). A similar calculations done to compute the three other joint probability $p(-1 + 1|\theta \theta{'})$, $p(+1 - 1| \theta \theta{'})$ and $p(-1 - 1| \theta \theta{'})$ where the result $-1$ is attributed to events where there is either \{ a
click in $a$ and no click in $a_\perp$ \} or \{ a click in $a$ and a click
in $a_\perp$\} (similarly for the modes $b^{\theta{'}}$ and $b^{\theta{'}}_\perp$).  We find
\begin{multline}
p(+1 + 1| \theta \theta{'}) = 
\\ =(1 - p_\text{dc}) \frac{1 - T^2_g}{1-(RT_g)^2}  \left[ f(\bar{n}, \bar{m}, \zeta ) - g(\bar{n}, \bar{m}, \zeta) \right] -
\\ -(1 - p_\text{dc})^2 \frac{(1 - T^2_g)^2}{(1-(RT_g)^2)^2}  \left[ f(\bar{m}, \bar{m}, \zeta) - g(\bar{m}, \bar{m}, \zeta) \right],
\label{eq:app_9}
\end{multline}
\begin{multline}
p(+1 - 1| \theta \theta{'}) =  \left[ f(\bar{n}, \bar{n}, \zeta ) - g(\bar{n}, \bar{n}, \zeta) \right] -
\\ -(1 - p_\text{dc}) \frac{(1 - T^2_g)}{(1-(RT_g)^2)}  \left[ f(\bar{n}, \bar{m}, \zeta) - g(\bar{n}, \bar{m}, \zeta) \right],
\label{eq:app_9_1}
\end{multline}
\begin{multline}
p(-1 + 1| \theta \theta{'}) = 
\\ = (1 - p_\text{dc}) \frac{1 - T^2_g}{1-(RT_g)^2} \left[ 1-f(\bar{n}, \bar{m}, \zeta ) \right]-
\\-(1 - p_\text{dc})^2 \frac{(1 - T^2_g)^2}{(1-(RT_g)^2)^2}  \left[ 1 -f(\bar{m}, \bar{m}, \zeta) \right],
\label{eq:app_9_2}
\end{multline}
\begin{multline}
p(-1 - 1| \theta \theta{'}) = \left[ 1-f(\bar{n}, \bar{n}, \zeta ) \right] -
\\ -(1 - p_\text{dc}) \frac{(1 - T^2_g)}{(1-(RT_g)^2)}  \left[ 1 -f(\bar{n}, \bar{m}, \zeta) \right] .
\label{eq:app_9_3}
\end{multline}
Here 
\begin{multline}
f(\bar{n}, \bar{m}, \zeta ) = \frac{1}{1+\cos^2\theta \bar{n}\eta + \sin^2 \theta \bar{m} \eta} \left( \frac{1}{1+4\zeta \epsilon} \right)^{1/2},
\label{eq:app_10}
\end{multline}
and
\begin{multline}
g(\bar{n}, \bar{m}, \zeta ) = \frac{1}{(1+ \bar{n}\eta)(1+ +  \bar{m} \eta)} \left( \frac{1}{1+4\zeta \epsilon} \right)^{1/2}
\label{eq:app_11}
\end{multline}
with
\begin{equation}
\zeta = \frac{T_2^2 \gamma^2 \eta_d \cos^2(\theta - \theta{'})}{1 + \cos^2 \theta \bar{n}\eta + \sin^2 \theta \bar{m}\eta},
\label{eq:app_12}
\end{equation}
\begin{equation}
\bar{\zeta} = T_2^2 \gamma^2 \eta_d \left( \frac{\cos^2 \theta{'}}{1+\bar{n}\eta} + \frac{\sin^2 \theta{'}}{1+\bar{m}\eta}\right)
\end{equation}
and
\begin{equation}
\eta = \eta_d T_1^2 T_2^2 \eta_c\eta_d,
\label{eq:app_13}
\end{equation}
\begin{equation}
\bar{n} = \frac{T_g^2}{1-T_g^2}, \bar{m} = \frac{(RT_g)^2}{1-(RT_g)^2}
\label{eq:app_14}
\end{equation}
and $\epsilon = 1 - V$ stands for the error on the viability associated to the displacement operation.  Note that the experiment is being performed under the fair sampling assumption, the four probabilities $p(\pm 1 \pm 1| \theta \theta{'})$,
$p(\pm 1 \mp 1| \theta \theta{'})$ are re-normalized to sum up to one before being used to predict the value of the CHSH inequality.


\end{document}